\documentclass[12pt]{article}

\usepackage[utf8]{inputenc}
\usepackage{charter}
\usepackage{fullpage}
\usepackage{hyperref}
\usepackage{graphicx}
\usepackage{lipsum}
\usepackage{hyperref}
\hypersetup{hidelinks}
\usepackage[total={6.5in,9in},top=1in,headsep=0.1in,headheight=1in]{geometry}
\usepackage{cite}

\newcommand{\fv}{f_\chi(\mathbf{v})}
\newcommand{\rhoDM}{\rho_\chi}

\newcommand\snowmass{
\begin{center}
  \rule[-0.2in]{\hsize}{0.01in}\\
  \rule{\hsize}{0.01in}\\
  \vskip 0.1in
  Submitted to the Proceedings of the US Community Study\\ 
  on the Future of Particle Physics (Snowmass 2021)\\
  \rule{\hsize}{0.01in}\\
  \rule[+0.2in]{\hsize}{0.01in}\\[-2em]
\end{center}
}

\usepackage[firstpage=true]{background}
\backgroundsetup{contents={\parbox{6.5in}{\snowmass}}, scale=1,placement=top,opacity=1,color=black,position={3.25in,1.2in}}

\usepackage{fancyhdr}
\fancypagestyle{plain}{%
  \fancyhf{}%
  \fancyhead[C]{}
  \fancyfoot[C]{\thepage}
}

\fancypagestyle{empty}{%
  \fancyhf{}%
  \fancyhead[C]{{\it Snowmass2021 Cosmic Frontier}}
  \fancyfoot[C]{\thepage}
}
\title{Snowmass2021 Cosmic Frontier: Modeling, statistics, simulations, and computing needs for direct dark matter detection}
\date{}

\usepackage{authblk}

\author[1,2,*]{Editors: Yonatan Kahn}
\author[3,4,5,*]{Maria Elena Monzani}
\author[6,*]{Kimberly J. Palladino}

\author[3,4]{\\Contributors: Tyler Anderson}
\author[7]{Deborah Bard}
\author[8]{Daniel Baxter}
\author[3,4]{Micah Buuck}
\author[3,4]{Concetta Cartaro}
\author[9]{Juan I. Collar}
\author[10]{Miriam Diamond}
\author[3,4]{Alden Fan}
\author[7,11]{Simon Knapen}
\author[7]{Scott Kravitz}
\author[12]{Rafael F. Lang}
\author[7]{Benjamin Nachman}
\author[13,7]{Ibles Olcina Samblas}
\author[14]{Igor Ostrovskiy}
\author[15]{Aditya Parikh}
\author[7]{Quentin Riffard}
\author[16]{Amy Roberts}
\author[8]{Kelly Stifter}
\author[17]{Matthew Szydagis}
\author[18]{Christopher Tunnell}
\author[19,20]{Belina von Krosigk}
\author[3]{Dennis Wright}
\author[21]{Tien-Tien Yu}
\author[3,4]{\\Endorsers: Dan Akerib}
\author[22]{Ray Bunker}
\author[23]{Thomas Y. Chen}
\author[24]{Graciela B. Gelmini}
\author[25]{Doojin Kim}
\author[26]{Jong-Chul Park}
\author[27]{Tarek Saab}
\author[28]{Rajeev Singh}
\author[29]{Shufang Su}
\author[30]{Yu-Dai Tsai}
\author[31]{Shawn Westerdale}
 
\affil[1]{Department of Physics, University of Illinois at Urbana-Champaign, Urbana, IL 61801, USA}
\affil[2]{Illinois Center for Advanced Studies of the Universe, University of Illinois at Urbana-Champaign, Urbana, IL 61801, USA}
\affil[3]{SLAC National Accelerator Laboratory, Menlo Park, CA 94025, USA}
\affil[4]{Kavli Institute for Particle Astrophysics and Cosmology, Stanford University, Stanford, CA 94305 USA}
\affil[5]{Vatican Observatory, Castel Gandolfo, V-00120, Vatican City State}
\affil[6]{Clarendon Laboratory, Department of Physics, University of Oxford, Parks Road, Oxford, OX1 3PU, UK}
\affil[7]{Lawrence Berkeley National Laboratory, Berkeley, CA 94720 USA}
\affil[8]{Fermi National Accelerator Laboratory (FNAL), Batavia, IL 60510-5011, USA}
\affil[9]{Enrico Fermi Institute, University of Chicago, Chicago, Illinois 60637, USA}
\affil[10]{Department of Physics, University of Toronto, Toronto, ON M5S 1A7, Canada}
\affil[11]{CERN, Theory Division, CH-1211 Geneva 23, Switzerland}
\affil[12]{Department of Physics and Astronomy, Purdue University, West Lafayette, IN 47907, USA}
\affil[13]{Department of Physics, University of California, Berkeley, CA 94720, USA}
\affil[14]{University of Alabama, Department of Physics and Astronomy, Tuscaloosa, AL 34587, USA}
\affil[15]{Department of Physics, Harvard University, Cambridge, MA 02138, USA}
\affil[16]{Department of Physics, University of Colorado Denver, Denver, CO 80217, USA}
\affil[17]{Department of Physics, The University at Albany, SUNY, Albany, NY 12222, USA}
\affil[18]{Department of Physics and Astronomy, Rice University, Houston, TX 77005, USA}
\affil[19]{Institute for Astroparticle Physics (IAP), Karlsruhe Institute of Technology (KIT), 76344, Germany}
\affil[20]{Institut f\"ur Experimentalphysik, Universit\"at Hamburg, 22761 Hamburg, Germany}
\affil[21]{Department of Physics and Institute for Fundamental Science, University of Oregon, Eugene,
Oregon 97403, USA}
\affil[22]{Pacific Northwest National Laboratory, Richland, WA 99352, USA}
\affil[23]{Fu Foundation School of Engineering and Applied Science, Columbia University, New York, NY 10027, USA}
\affil[24]{Department of Physics and Astronomy, University of California, Los Angeles, CA 90095, USA}
\affil[25]{Department of Physics and Astronomy, Texas A\&M University, College Station, TX 77843, USA}
\affil[26]{Department of Physics, Chungnam National University, Daejeon 34134, Republic of Korea}
\affil[27]{Department of Physics, University of Florida, Gainesville, FL 32611, USA}
\affil[28]{Institute of Nuclear  Physics, Polish Academy of Sciences,  PL-31-342 Krak\'ow, Poland}
\affil[29]{Department of Physics, University of Arizona, Tucson, AZ 85721, USA}
\affil[30]{Department of Physics and Astronomy, University of California, Irvine, CA 92697-4575, USA}
\affil[31]{Department of Physics and Astronomy, University of California, Riverside, CA 92521, USA}
\affil[*]{Editors: \href{mailto: yfkahn@illinois.edu}{yfkahn@illinois.edu}, \href{mailto: monzani@stanford.edu}{monzani@stanford.edu}, \href{mailto: kimberly.palladino@physics.ox.ac.uk}{kimberly.palladino@physics.ox.ac.uk}}

\begin{document}

\maketitle
\noindent

\begin{abstract}

This paper summarizes the modeling, statistics, simulation, and computing needs of direct dark matter detection experiments in the next decade.

\end{abstract}

\section{Modeling}

In this section we describe our evaluation of the DM direct detection community's needs for modeling, which we define as theoretical input needed to better understand the expected DM signal at current detectors, as well as input which could help determine figures of merit for future detectors. As the next generation of experiments approaches important milestones in parameter space, the goal of such modeling should be to allow a robust claim of exclusion or discovery at a given level of statistical significance; this is especially important for new small-scale experiments searching for sub-GeV DM, where many-body effects in the target material can affect the signal model in important ways. On the other hand, the modeling needs for the DM phase space distribution (local density and velocity profile) are common to both sub-GeV and WIMP experiments. We summarize these needs here, in rough order of priority:
\begin{itemize}
    \item Framework for updating choices of $\rho_\chi$ and $f_\chi(\mathbf{v})$ which incorporate the best current astrophysical knowledge, rather than just a standard benchmark which may be based on outdated parameters (WIMP DM and sub-GeV DM)
    \item Data-driven detector response functions including many-body effects (sub-GeV DM)
    \item Many-body response functions for DM-SM interactions beyond the spin-independent benchmark (sub-GeV DM)
    \item Compendium of material properties for optimal DM detectors which could facilitate selection of future detector material (sub-GeV DM)
\end{itemize}


\subsection{Modeling the DM phase space distribution}

\begin{figure}[t!]
 \includegraphics[width=0.45\textwidth]{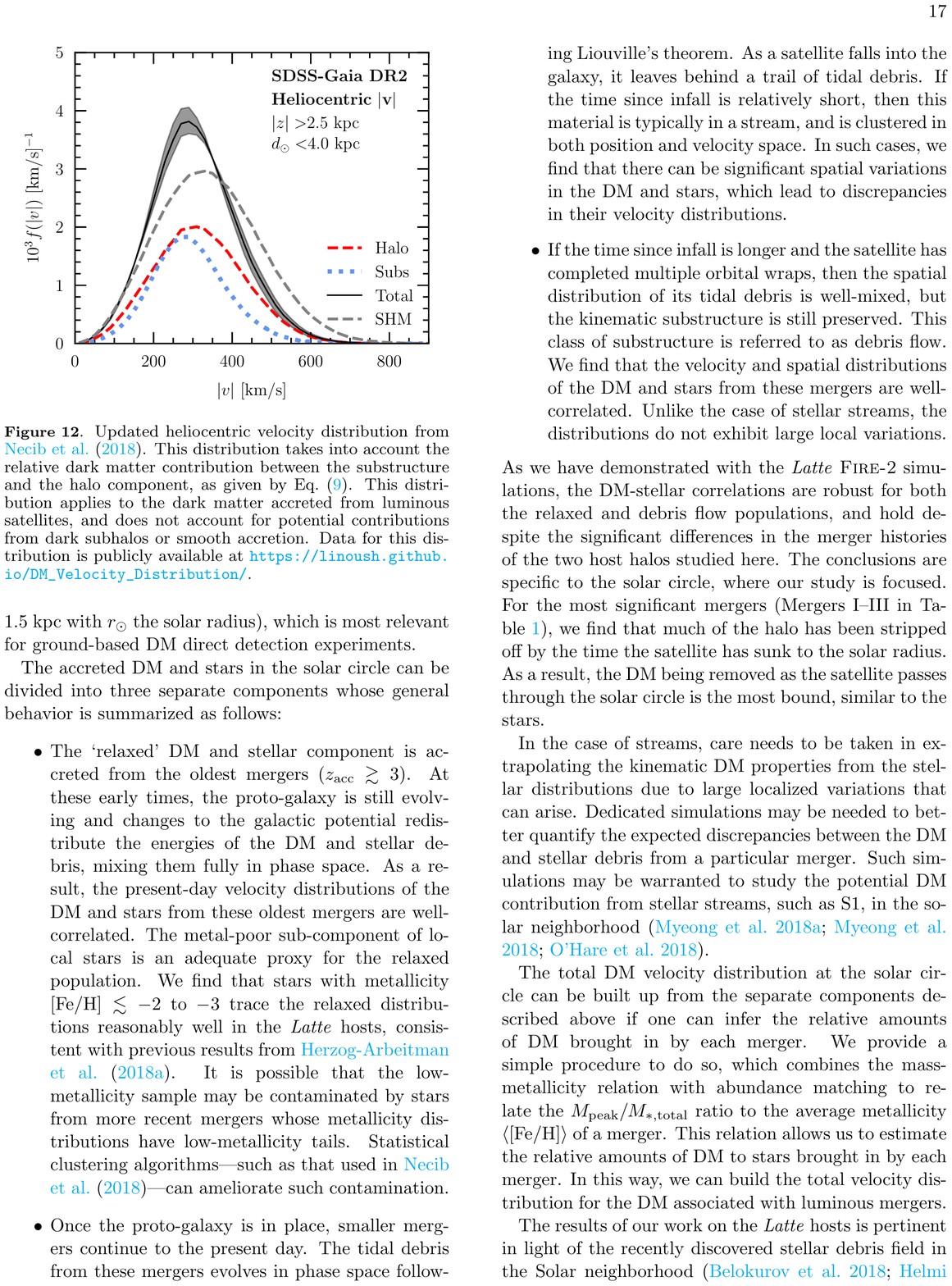} \qquad
  \includegraphics[width=0.45\textwidth]{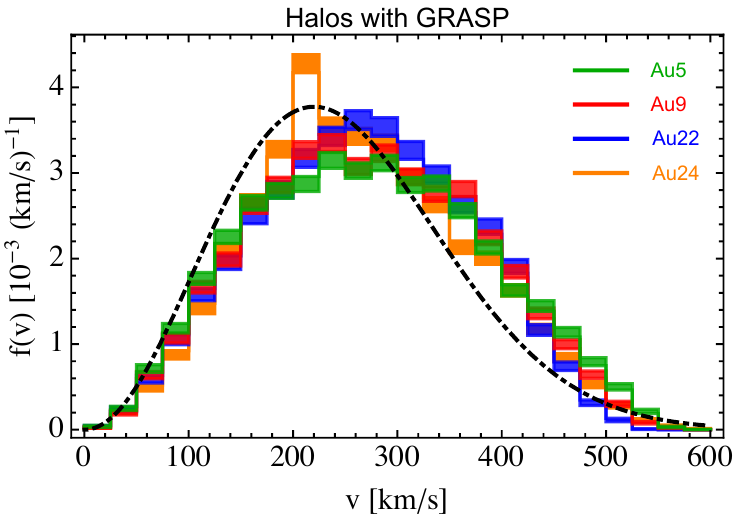} 
\caption{\label{fig:DMHalo} Non-Maxwellian components of the DM velocity distribution as extracted from N-body simulations: \texttt{Fire-2} \cite{Necib:2018igl} (left) and Auriga \cite{Bozorgnia:2019mjk} (right). ``GRASP'' refers to the radially-anisotropic stellar population identified by the \emph{Gaia} survey.}
\end{figure}

Converting the result of a direct detection experiment (a null result, or in the most optimistic case, events in the signal region) into a cross section limit or claim of discovery requires assumptions about the lab-frame DM velocity distribution $\fv$ and the local DM density $\rhoDM$. Since the last Snowmass study, there has been substantial progress in improving our knowledge of these parameters which are necessary inputs for any signal model, including studies combining precision astrometric data (especially from \emph{Gaia} DR2~\cite{gaia_collaboration_gaia_2018}) with N-body simulations (see for example Refs.~\cite{Herzog-Arbeitman:2017fte,Herzog-Arbeitman:2017zbm,Necib:2018iwb,Evans:2018bqy,Necib:2018igl,Bozorgnia:2018pfa,Bozorgnia:2019mjk,Besla:2019xbx,OHare:2019qxc,deSalas:2020hbh,Necib:2021vxr}, as well as Ref.~\cite{Benito:2020lgu} which infers the DM distribution from rotation curves assuming a model for the density profile). There is now persuasive evidence that some fraction of the local DM distribution differs from the Maxwellian Standard Halo Model (SHM), as shown in Fig.~\ref{fig:DMHalo}. As reviewed in Ref.~\cite{Baxter:2021pqo}, in practice, these deviations only affect the analysis very close to threshold~\cite{DEAP:2020iwi}, and changes in $\rhoDM$ are simply overall rescalings of the resulting exclusion limits. On the other hand, in the next decade DM experiments will begin to probe parameter space with very sharp theory predictions for cross sections, for example the freeze-out and freeze-in targets~\cite{Battaglieri:2017aum}. A small change in the DM distribution could meaningfully affect whether regions of parameter space are considered to be ruled out at high statistical significance. 

While we are sympathetic to the need for a common benchmark DM distribution model that can be used across experiments, we should not discount the work that is being done on the astrophysics side to better measure and model the observed distribution. It would therefore be of strong interest to build an interactive community-maintained database of various parameterizations of $\fv$ and $\rhoDM$ beyond the SHM, which could potentially be used by experimental collaborations to show the range of systematic uncertainty on exclusion limits. Alternatively, halo-independent methods have also been developed, which can avoid the uncertainties introduced by such models and which apply to both DM-nuclear and DM-electron scattering ~\cite{Drees:2007hr,Fox:2010bz,Chen:2021qao}. Indeed, both methods (astrophysically-informed halo models and halo-independent) are complementary for the analysis of putative signals, and we encourage both the theoretical and experimental communities to develop both analysis methods to the extent possible.

\subsection{Modeling the DM-target interaction for existing detectors}
\label{sec:DetResponse}

The signal modeling in sub-GeV DM experiments (for example, a predicted quantized charge spectrum from a CCD detector, or a heat spectrum from a phonon detector) is qualitatively different from traditional WIMP experiments because there is no strong separation of scales between the DM kinematics and the internal degrees of freedom of the target; while nuclear recoils for a 100 GeV WIMP can be modeled to an excellent approximation as free-particle 2-to-2 scattering~\cite{Lewin:1995rx}, the same is not true for sub-GeV DM which primarily interacts with quasiparticles or collective modes~\cite{Kahn:2021ttr}. Accurate modeling of the DM signal in these detectors will necessarily involve close collaboration with condensed matter physicists and materials scientists. An important tool which has gained prominence in the community since the last Snowmass study is density functional theory (DFT), which can predict electron and phonon band structures for a wide variety of materials with knowledge of the lattice structure. The DM-electron signal may then be computed using a number of public code packages recently developed by theorists (both DM and condensed matter): either in terms of single-particle wavefunctions extracted from DFT, as in \texttt{QEDark} \cite{Essig:2015cda}, \texttt{QEDark-EFT} \cite{Catena:2021qsr}, or \texttt{EXCEED-DM} \cite{Griffin:2021znd}; or in terms of the many-body response function known as the energy loss function (ELF) \cite{Hochberg:2021pkt,Knapen:2021run}, as in \texttt{DarkELF} \cite{Knapen:2021bwg}.\footnote{See also \texttt{DarkARC} \cite{Catena:2019gfa} for computing response function in atomic targets such as Ar and Xe, and \texttt{PhonoDark} \cite{Trickle:2020oki} for computing single-phonon production rates in diverse crystal targets.} Currently-operating experiments are not yet at single-phonon threshold, but as heat detectors achieve thresholds below the ion displacement energy (about 20--40 eV in silicon), modeling the signal in terms of the multi-phonon response will be necessary \cite{Kahn:2020fef,Knapen:2020aky}. 

\begin{figure}[t!]
 \includegraphics[width=0.45\textwidth]{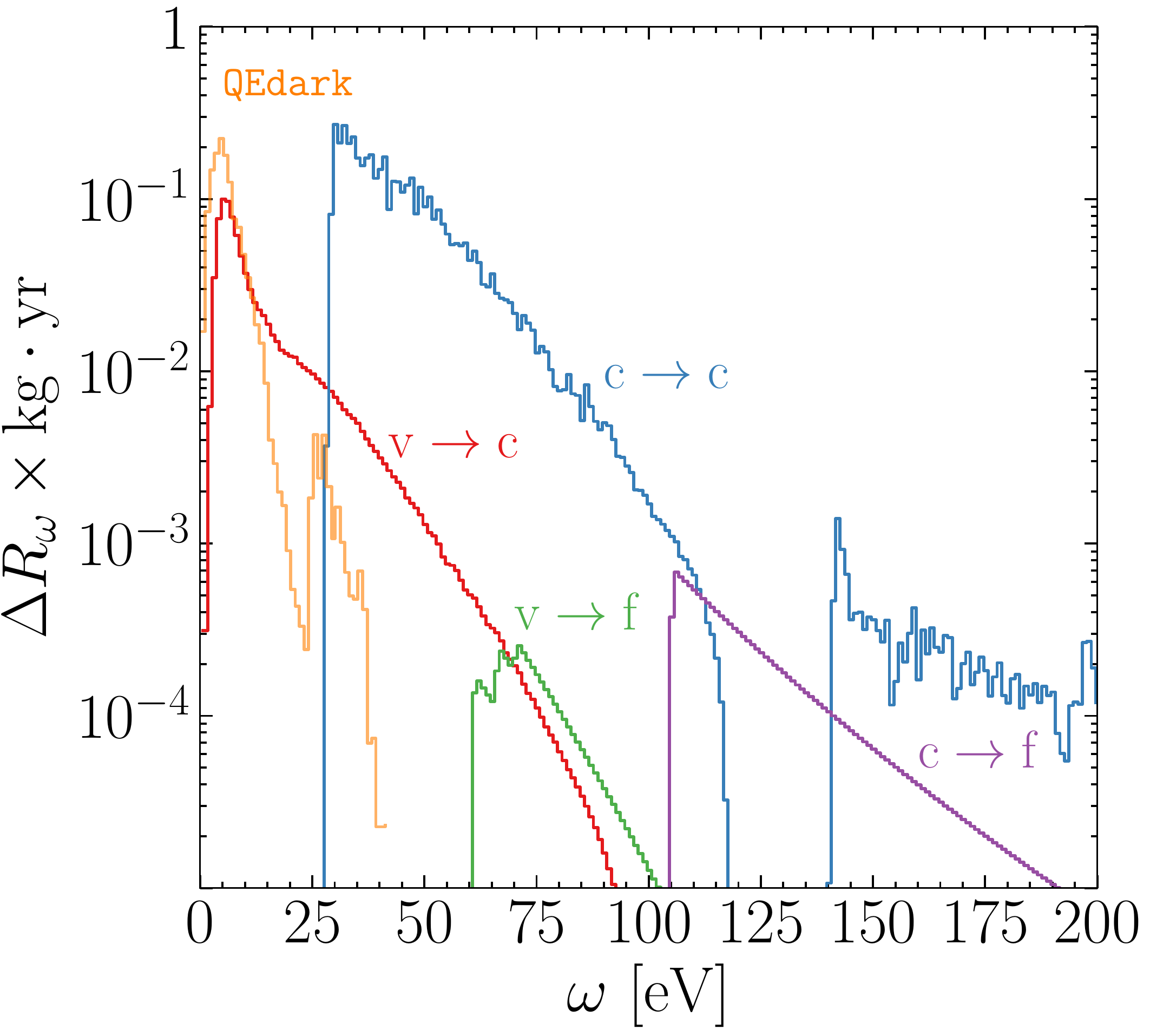} \qquad
  \includegraphics[width=0.45\textwidth]{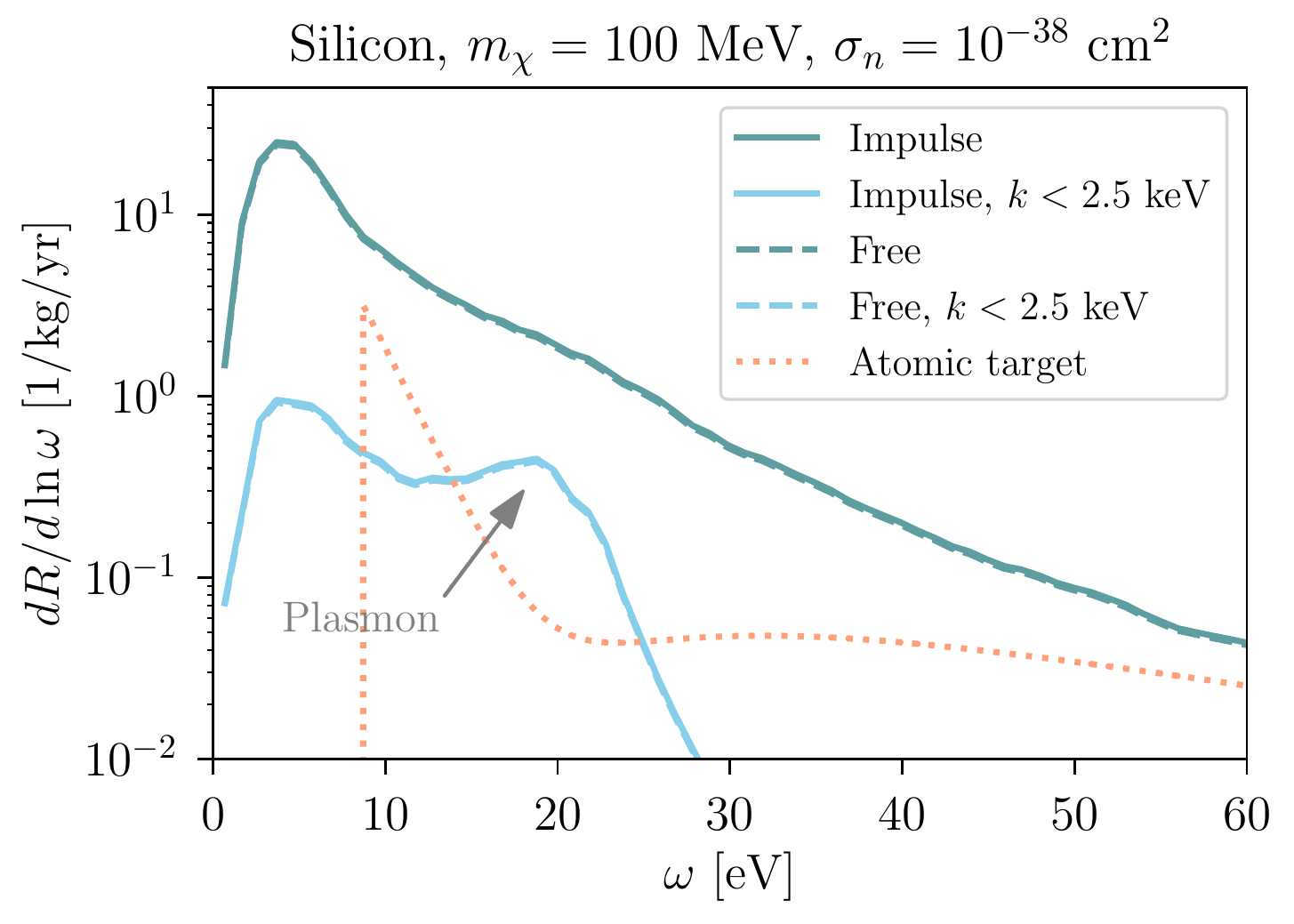} 
\caption{\label{fig:DetectorModeling} Different models of the detector response can change the predicted signal spectrum by orders of magnitude, as shown for electron scattering in germanium \cite{Griffin:2021znd} (left) and the Migdal effect in silicon (right) \cite{Knapen:2020aky}.}
\end{figure}

Since DM occupies an unusual kinematic regime not typically probed by condensed matter experiments, the majority of these signal models are at this time experimentally unvalidated, and different model choices can lead to quite prominent changes in the spectrum and rate. For example, all-electron reconstruction~\cite{Griffin:2021znd} predicts a peak at 25 eV in Ge which is not present in the plane-wave basis used in \texttt{QEDark} (Fig.~\ref{fig:DetectorModeling}, left); there are orders of magnitude differences in the rate of Migdal ionization between the \texttt{QEDark} treatment~\cite{Essig:2019xkx} and one based on the ELF which accounts for the preferred reference frame of the lattice~\cite{Knapen:2020aky,Knapen:2021bwg} (Fig.~\ref{fig:DetectorModeling}, right); and the rate of electronic excitation from DM absorption depends sensitively on both the spin/parity of the DM and many-body effects in the target \cite{Gelmini:2020xir,Mitridate:2021ctr}. Calibrating these response functions is a primary goal of the Cosmic Frontier particle-like DM topical group \cite{SnowmassCF1WP3}, and is an essential need for the sub-GeV DM community over the next decade, but even interpreting calibration data from techniques such as electron energy-loss spectroscopy (EELS) may require significant theoretical condensed matter physics to extract the ELF. In the meantime, we recommend that experimental collaborations consider including a variety of signal models from these various treatments of the response function in order to give an indication of the magnitude of the systematic uncertainty. The recently-released package \texttt{obscura} \cite{Emken:2021uzb} aims to facilitate this process.

\subsection{Modeling the fundamental DM-SM interaction}

The most common benchmark models used in sub-GeV direct detection assume a dark photon mediator, which yields a spin-independent coupling to charge density in an arbitrary target. However, to cover as much parameter space as possible with a minimum of theoretical bias, the community should also develop tools to probe all possible DM-electron or DM-nucleon operators allowed in an effective field theory (EFT) framework. The enumeration of these operators was completed in Refs.~\cite{Catena:2019gfa,Trickle:2020oki,Catena:2021qsr} for both isolated atomic targets and crystals, with Ref.~\cite{Trickle:2020oki} focusing on excitation of collective modes such as phonons and magnons, and Ref.~\cite{Catena:2021qsr} focusing on electronic excitation (see also Refs.~\cite{Liu:2021avx,Chen:2022pyd} for spin-dependent DM-electron interactions). As all of these works have noted, many of these EFT-derived response functions are novel because they involve interactions not present in the Standard Model (SM), which also makes them quite difficult to measure. In the spirit of data-driven calibration, it would be useful to know which of these operators can in principle be extracted with SM probes such as neutrons, photons, electrons, or electromagnetic fields. Indeed, there are already important differences between the spin-independent response derived from DFT wavefunctions compared to the ELF (in particular, the plasmon \cite{Kurinsky:2020dpb,Kozaczuk:2020uzb}), and thus one should view these novel response functions as subject to experimental verification. For example, a coupling between probe velocity and electron spin, analogous to the spin-orbit coupling which contributes to the fine structure of hydrogen, is generated both by DM interacting via a light dark photon mediator and electrons interacting via ordinary electromagnetism, and thus may potentially be extracted from high-precision electron scattering experiments \cite{Kahn:2021ttr}. Simultaneously, further theoretical work is needed to determine which EFT models are compatible with a consistent thermal history which does not violate cosmological or astrophysical constraints, which can be quite severe even for weakly-coupled new physics in the MeV--GeV regime.

\subsection{Modeling the detector response for future detectors}
Many novel materials have been proposed as sub-GeV DM detectors (see for example Refs.~\cite{Guo:2013dt,Hochberg:2015fth,Hochberg:2016ntt,Cavoto:2017otc,Hochberg:2017wce,Griffin:2018bjn,Geilhufe:2018gry,Kurinsky:2019pgb,Geilhufe:2019ndy,Trickle:2019ovy,Griffin:2019mvc,Griffin:2020lgd,Hochberg:2021pkt,Coskuner:2021qxo,Blanco:2021hlm}), and there are now active collaborations devoted to developing detectors based on some of these materials, including superconductors \cite{Hochberg:2019cyy,Hochberg:2021yud}, anisotropic low-gap insulators \cite{LOI_CF1_CF2-IF1_IF2_Kurinsky-029}, superfluid helium-4 \cite{SPICEHeRALD:2021jba}, polar crystals \cite{LOI_SNOWMASS21-CF1_CF2-IF1_IF8-120}, and graphene Josephson junctions~\cite{Kim:2020bwm,LOI_SNOWMASS21-CF1_CF0-TF9_TF0-IF2_IF0_Doojin_Kim-026} (see also \cite{SnowmassCF1WP2} for a review of recent progress). Given the investment of time, money, and person-power required to develop a new detector, it will be worthwhile for the community to have some indication of the figures of merit which motivate the use of one detector material over another. Some initial work along these lines has recently taken place. Ref.~\cite{Griffin:2019mvc} identified a material-specific ``quality factor'' depending on the Born effective charges, atomic masses, optical phonon spectrum, and high-frequency dielectric constant which maximized the single-phonon production rate at large DM masses. Similarly, the relative size of the DM velocity compared to a sound speed \cite{Coskuner:2021qxo} or an effective electron velocity \cite{Hochberg:2017wce,Hochberg:2021pkt,Blanco:2021hlm} controls both the total rate and the daily modulation amplitude in anisotropic materials. One can also use first-principles constraints from ``sum rules'' on many-body response functions to identify optimal response functions which theoretically maximize the total rate \cite{Lasenby:2021wsc}, and from there attempt to identify realistic materials which best approximate them. Since all of these analyses make different assumptions about the DM interactions and the modeling of the target response, it would be useful to synthesize this information in such a way as to facilitate rapid high-throughput searches from existing catalogues of material properties to identify new candidate detector materials \cite{Jain2013,Geilhufe:2018gry,Inzani:2020szg}. As much of this work is on the leading edge of current research, these studies may arise organically from the community over the next decade.

\section{Statistics and Analysis}

With the sophistication of modern direct dark matter experiments, and the varieties of models to be explored, the statistical techniques employed have similarly matured. Currently, methods based on the Profile Likelihood Ratio (PLR) are the most commonly employed in searches for weakly-interactive dark matter candidates. However, the specific choice of test statistic has fluctuated over the years, even within the same collaboration~\cite{lux_first,lux_collaboration_results_2017,Aprile:2017iyp,xenon1t_sr1}.

This was one of the main topics that was addressed at the 2019 PHYSTAT-DM workshop. Participants from DAMIC, DarkSide, DARWIN, DEAP, LZ, NEWS-G, PandaX, PICO, SBC, SENSEI, SuperCDMS, and XENON collaborations achieved a consensus on a number of important topics, mostly centered around the PLR method but not exclusively. The results of this work were summarized in Ref.~\cite{Baxter:2021pqo}. We endorse the common standards presented in that reference, which, if they are widely adopted in the direct detection community, will facilitate the comparison of statistical claims amongst different collaborations. In particular, and mostly following the conventions introduced in that work, we recommend:

\begin{itemize}
    \item For standard WIMP searches, we recommend conducting a PLR analysis on a per-mass basis and using the test statistic in Eq.~(12) of Ref.\cite{Cowan:2010js}. Furthermore, claims of evidence should require at least a 3$\sigma$ global discrepancy with the background-only hypothesis.
    \item Upper limits should be based on a two-sided PLR test statistic in which the parameter of interest is bounded from below to 0 (Eq.~(11) of Ref.~\cite{Cowan:2010js}). To continue with past conventions, any limit or projection should be presented at the 90\% confidence level (CL).
    \item To avoid the exclusion of parameter space to which an experiment has a vanishingly small sensitivity, we recommend the use power-constrained limits (PCL), with a critical power of 0.32 (see Ref.~\cite{PCL} for further information).
    \item We urge collaborations to make their data more usable by the larger physics community. For instance, by sharing the data points of exclusion limits or opening their statistical models/likelihoods for cross-examination.
    \item In order to mitigate possible biases introduced while analysing data, collaborations should follow blinding or salting procedures to the extent possible. Section~2.6 of Ref.~\cite{Baxter:2021pqo} provides more information on this topic.
    \end{itemize}

\noindent In addition, there are statistical considerations which are specifically important for sub-GeV DM searches. As this field is only a decade old and rapidly-developing (see Ref.~\cite{SnowmassCF1WP2}), we frame these considerations as needs for the future rather than recommendations for current implementation:
\begin{itemize}
    \item A number of proposals for future sub-GeV DM experiments predict large daily modulation amplitudes~\cite{Hochberg:2016ntt,Budnik:2017sbu,Cavoto:2017otc,Griffin:2018bjn,Coskuner:2019odd,Geilhufe:2019ndy,Blanco:2019lrf,Blanco:2021hlm,Coskuner:2021qxo}, typically around $10\%$ for a wide range of masses and as large as $\mathcal{O}(1)$ near threshold. There have been several proposals in the literature for test statistics relevant to daily modulation~\cite{Geilhufe:2019ndy,Blanco:2021hlm,Hochberg:2021ymx}; the simplest is a 2-bin analysis, but with the large event rates possible at light DM experiments, an analysis in terms of harmonics may be possible~\cite{Lee:2013xxa}, as well as a joint daily modulation-annual modulation analysis for long-exposure runs. A common statistical framework for these modulation effects would be especially useful, both for discovery and for rejecting unknown (flat) backgrounds.
    \item As experiments push thresholds lower and lower, new backgrounds will inevitably appear. For example, the source of the persistent rising low-energy spectra in eV-threshold calorimetric experiments is still being actively discussed~\cite{Proceedings:2022hmu}. It would be useful for the community to explore statistical tests for systematic mismodeling which are agnostic as to the physical origin of the background; such techniques are common in astrophysics, and for example have recently been applied to direct detection experiments~\cite{Dessert:2020vxy}. These, and similar issues, are being pursued as a follow-up to the PHYSTAT-DM report, this one focused specifically on sub-GeV DM.
\end{itemize}

\section{Simulations}
\label{sec:Simulations}

Simulations are an essential component of modern physics experiments. They are vital at every stage of the detector life-cycle, starting with the initial design phase, all the way to final data analysis. In some cases, the associated costs to support computing and code development may represent a large fraction of the experiment's budget. Therefore, it is essential to have an efficient, well-maintained, well-understood and thoroughly validated simulation infrastructure. A full-chain simulation infrastructure includes both event generators and detector simulation frameworks. The common needs to all direct dark matter detection experiments include:

\begin{itemize}
    \item The continuation of Geant4~\cite{ALLISON2016186} support and training within the community.
    \item Continued support for event generators, including those developed as part of a national security program.
    \item Detector-specific simulation packages, such as: NEST~\cite{Szydagis_2011,Mock_2014}, which simulates noble elements detector response, and Opticks~\cite{Simon_blyth_Opticks}, which tracks optical photons.
    \item Opportunities for cross-collaboration communication, in order to reduce duplication of effort and maximize return on investment.
\end{itemize}

\noindent For low-threshold (sub-keV) experiments in particular, a further essential need is:
\begin{itemize}
    \item Consistent incorporation into existing codes of solid-state effects such as directionally-dependent displacement energy, modifications to the Lindhard ionization yield model, and realistic atomic binding energies.
\end{itemize}

\subsection{Common Frameworks}
\label{subsec:SimFrameworks}

The Geant4 toolkit~\cite{ALLISON2016186} underlies most of the simulation frameworks utilized in dark matter detection. However, U.S. funding for this project was discontinued in recent years. As a result, the community can no longer rely on core Geant4 developers to address the needs of neutrino and DM experiments. This represents a three-fold challenge: no further updates to the Geant4 engine are expected; the physics lists underlying the different models of Geant4 will not be updated as often; and finally, user training and support for U.S.-based adopters, which was provided by core Geant4 developers, has all but disappeared. Continued support for Geant4 is crucial to the design and construction of future experiments, and for the interpretation of their results.

Commonality of software within the community is highly desirable, as it increases reliability, reduces duplication of effort, and eases the maintenance burden. Several components would aid in this: a shared software repository and continuous integration facility, a core Monte Carlo engine, a collection of physics databases from which the simulation can draw, and a set of event generators that are commonly used by the various experiments. In the current model, each collaboration builds a custom Monte Carlo framework based on Geant4. However, much of this effort shares deep similarities across experiments. The establishment of a common framework geared towards the need of the direct detection community would encourage the sharing of new algorithms and physics models. It would also facilitate the validation of the physics output using data from multiple detectors.

\subsection{Detector Simulations}

Each class of detectors requires the development of specialized code, to deal with the detailed response to very low energy recoils. In the case of noble liquids, two of the  most computing-intensive challenges are: the simulation of scintillation light with its thousands of optical photons, and the drifting of thermal electrons. The large number of particles to be tracked in these applications is a simulation bottleneck: the passage of one charged particle will generate many low-energy electrons or photons, which in turn also need to be tracked, thus greatly increasing computation time.

Opticks may provide a solution to the tracking of optical photons. Opticks~\cite{Simon_blyth_Opticks} is a GPU-based ray-tracing code supporting an integration with the Geant4 toolkit. It is estimated to be up to 1,000 times faster than Geant4 alone for the tracking of optical photons. A similar code for thermal electrons does not yet exist, but one could be modeled after Opticks. Reducing the computational cost of a full detector simulation involving optical photons and thermal electrons will lead to a drastic improvement in the level of detail available to our detector model. It will also allow for a better understanding of backgrounds, which are a challenge in rare event searches.

A toolkit used by multiple collaborations is NEST~\cite{Szydagis_2011,Mock_2014}, which simulates the excitation, ionization, and scintillation processes in noble elements. NEST exists both as a standalone executable and as a callable Geant4 library. Integrating NEST with Opticks would fully leverage the GPU gains in optical photon simulations. A package like NEST is maximally beneficial to the community thanks to its modularity, the quality of its documentation, and the robust plans for long-term support. 

The response of detector electronics and acquisition chain also needs to be simulated, all the way to a DAQ-like format. This task requires generators for light and charge sensors, cables, and analog and digital electronics. This type of simulation tends to be developed in-house by each experiment (see for example~\cite{LUX-ZEPLIN:2020lib}). But as it is a common need in the field, an increase in code sharing is much desired. Analogous considerations apply to the needs for detector visualization: Geant4 provides a comprehensive set of interfaces and drivers~\cite{Allison:2007zzb}, but the customization of these interfaces to different experimental ecosystems tends to be quite burdensome. 

\subsection{Detector Response in the sub-keV regime}

As experiments progress to lower thresholds, low-energy physics must be consistently implemented in phenomenological codes such as SRIM~\cite{ZIEGLER20041027}, MCNP~\cite{werner2017mcnp,werner2018mcnp}, and Geant4. In the past decade, significant effort has been invested in improving the low-energy modeling of solid state detectors in Geant4 by modeling the propagation of acoustic phonons, electrons and holes in cryogenic crystals~\cite{brandt2014semiconductor}. This effort was spurred by the simulation needs of the SuperCDMS detectors~\cite{2014JLTP}, and the transport code was implemented directly in Geant4 to ensure availability to the wider scientific community.

At present, the two standard solid-state nuclear recoil codes, SRIM and MCNP,  are based on the Lindhard model of ionization yield \cite{lindhard1963integral}, for which nuclear recoils below 100 eV are explicitly outside the domain of validity of the model. There are also persuasive experimental hints that the Lindhard model fails for sub-keV recoils \cite{Chavarria:2016xsi,Izraelevitch:2017gfi,Collar:2021fcl,SuperCDMS:2022nlc}, along with some initial attempts to account for many-body effects, including more realistic atomic binding energies \cite{Sarkis:2020soy} and ab-initio modeling of the material response with time-dependent DFT \cite{lim2016electron,horsfield2016adiabatic}. The general Markov chain framework of both SRIM and MCNP, where the energy deposit to a single primary recoil is distributed among secondaries, may actually not represent the physical situation at low energies, which resembles more closely a cloud of secondary recoils carrying 10's of eV apiece. At energies below the ion displacement energy ($\sim 20-40$ eV in silicon), the picture of a primary recoil no longer makes sense as the energy is deposited directly into phonons and/or charge.

As theoretical modeling improves, informed by new low-threshold calibration experiments, the relevant physics may be clarified and incorporated into simulations. This is closely related to the detector modeling needs of Sec.~\ref{sec:DetResponse} above: Sec.~\ref{sec:DetResponse} focuses on modeling the primary DM-target interaction, while the simulations carry this process forward to predict the end-stage ionization or heat signals seen by the detector. 

\subsection{Event Generators and Physics Lists}

The Geant4 toolkit contains predefined physics lists that provide options for modeling various processes, intended to align with a specific application. The toolkit also contains the functionality to allow user-defined processes to be integrated into the physics of the simulation. These physics lists have been optimized and validated by several community modifications that improve their modeling accuracy. Keeping these physics lists up-to-date and validating their output is one of the main efforts in maintaining the framework.

The most accurate physics lists that at low energies (which makes them most relevant for WIMP direct detection), include: G4EMLivermorePhysics, which covers electromagnetic interactions using Livermore models for gamma and electron cross-sections~\cite{osti_295438,osti_5691165}, extending the validity of the physics down to 10 eV; and G4HadronPhysicsQGSP\_BIC\_HP, which applies the Binary Cascade (BIC) intra-nuclear model~\cite{Folger:908839} for lower energy inelastic interactions, and uses the the High Precision neutron models and cross sections for neutrons of 20 MeV and lower.

Geant4 provides functionality to simulate radioactive decays, using data libraries from the Evaluated Nuclear Structure Data File (ENSDF)~\cite{TULI1996506}, which describes the nuclear decays, and from the Livermore Evaluated Atomic Data Library (EADL)~\cite{osti_5691165}, which describes any subsequent atomic transitions. The modeling of nuclear reactions, including neutron yields and spectra from spontaneous fission, $(\alpha,n)$ and delayed neutron emission due to the decay of radionuclides can be obtained by interfacing with packages like TALYS~\cite{TALYS} and SOURCES4A~\cite{Sources4A}. Ensuring support for such a variety of custom tools is crucial for the success of the field in the upcoming decade.

\section{Computing}
\label{sec:Computing}

Current dark matter experiments are approaching data volumes of order 1 PB/year~\cite{Mount:2017qzi}. This is not an unusual scale for HEP, however it presents a significant challenge in the direct detection community, which until recently hasn't prioritized the development of a scalable computing infrastructure in support of its scientific ambition. Moreover, a fragmentation of funding sources and a climate of competition between experiments, have hindered the opportunities for cooperation and tool sharing in the computing and software domain, leading to unnecessary duplication of efforts across the field. Key strategic goals to ensure success for the experiments of the next decade include:

\begin{itemize}
    \item Lower the barrier of entry to national supercomputing facilities, by providing common tools and shared engineering. Solicit community input on architecture evolution, while providing access to specialized resources, such as GPU and TPU clusters.
    \item Support scalable software infrastructure tools across HEP, avoiding duplication of effort. These tools run the gamut of data management and archiving, event processing, reconstruction and analysis, software management, validation and distribution.
    \item Enhance industry collaborations on machine learning techniques and provide access to external experts. Foster community-wide efforts to understand uncertainties and physical interpretation of machine learning results.
    \item Expand training and career opportunities for computing-inclined physicists, both within academia and in cooperation with industry. Ensure that academic and national laboratory positions are viable career options for a diverse group of people.
    \end{itemize}

\subsection{Computing Model Evolution}
\label{subsection:CompModel}

Direct detection experiments are growing more reliant on High-Performance Computing (HPC) centers for a significant fraction of their computing needs. Leveraging HPC resources involves significant challenges for the experiments, in porting their frameworks to HPC resources, and for the HPC center staff, in accommodating an increasing number of users with limited HPC expertise. What distinguishes the direct detection community is that the common issues faced by all experiment teams are exacerbated by the lack of human resources available to work on these issues. It is vital that science teams and HPC centers partner to address these challenges, which fall into roughly four areas:  
\begin{itemize}
    \item Getting codes running on continuously evolving architectures
    \item Adapting to HPC center policy (security and policy optimization)
    \item Operational tension in prioritizing performance over 24/7 availability
    \item Provide (adopt) common tools that are easy to stand up and maintain
\end{itemize}

As described in sections~\ref{subsec:SimFrameworks} and~\ref{subsec:HepFrameworks}, HEP experiments rely on a set of frameworks designed in the pre-HPC era. These frameworks utilize a sizeable memory footprint compared to the average RAM/core available on supercomputers, are built as monolithic single-executable objects, and require such a level of customization that make it challenging to incorporate their latest features on a regular basis. In addition, the HPC architecture paradigms tend to change faster than the lifetime of each experiment (approximately twice per decade), and each collaboration is faced with the challenge of adapting their codebase to this continuously evolving model with insufficient resources.

Breaking up these frameworks into more modular components is desirable, but would require a massive validation effort of the physics output, which in turn would require a significant investment from the community. The NESAP program at NERSC (the National Energy Science Research Computing center) partners with application development teams and vendors to port and optimize codes to new architectures. 
Programs like NESAP are extremely valuable to the direct detection community, notwithstanding the recruiting and retention challenges described in~\ref{subsec:Workforce}.

Some of the biggest challenges faced by users of HPC centers are due to security and policy concerns, rather than technical barriers. Supporting the needs of experiment workflows has pushed some centers to change their policies around near-realtime access to systems, and support for federated ID. NERSC has recognised that running experiments highly value continued access to its services (for example to monitor detector status 24/7), and is now able to keep a subset of its infrastructure operational during outages on generator power; moreover, they invested in new system software capabilities for rolling upgrades. Similar improvements are desirable at all HPC centers supporting HEP, together with developing and supporting portable cross-facility workflows\cite{alcc}.

The LBNL Superfacility Project~\cite{superfacility} was designed to leverage and integrate work being done across NERSC, ESnet and research divisions at LBNL to provide a coordinated and coherent approach to supporting experiments at DOE facilities. The project aims to provide an integrated, scalable and sustainable framework for experiment science, working closely with a range of science teams on design and requirements, and using industry standard and open source tools wherever possible. NERSC is now able to support automated pipelines that analyze data from remote facilities at large scale, using capabilities such as near-realtime computing, dynamic networking, API-driven automation, HPC-scale notebooks with Jupyter~\cite{jupyter}, state of the art data management tools, federated ID and container-based peripheral services. Similar initiatives at other HPC facilities would be extremely beneficial to the community.

\subsection{Scalable HEP frameworks}
\label{subsec:HepFrameworks}

A distinguishing feature of direct detection experiments, compared for example to experiments at colliders, is the lack of a significant investment in software and computing. This causes an inability to develop dedicated tools for data acquisition, simulation, data handling, and reconstruction tools, which are often developed from scratch by large collider experiments.  In some cases, LHC experiments have (at least partially) abstracted and open-sourced their software in a way that it can be re-used by others. In addition to the simulation frameworks mentioned in~\ref{sec:Simulations}, these tools can be categorized as following:

\begin {itemize}
    \item Data management and archiving (Rucio~\cite{Barisits:2019fyl}, DIRAC~\cite{Tsaregorodtsev:2014kda}, globus~\cite{Globus:2006})
    \item Event processing, reconstruction and analysis (Gaudi~\cite{Barrand:2001ny}, Art~\cite{Green:2012gv}, LArSoft~\cite{Snider:2017wjd}, Acts~\cite{Ai:2019kze}, ROOT~\cite{Brun:1997pa})
    \item Software management, validation and distribution (GitLab/GitHub, easybuild~\cite{Villanueva2019-qs}, LCGCMake~\cite{Villanueva2019-qs}, CVMFS~\cite{Timm:2017mtp}, Coverity~\cite{coverity:2016}, Singularity~\cite{singularity:2017}, Shifter~\cite{shifter:2017})
\end{itemize}

The adoption of open source tools by the direct detection community requires significant development and adaptation. One overarching theme is that LHC-driven tools feature a strict event-based organization, because their detectors are triggered in coincidence with a collider interaction. However, dark matter detectors observe a continuous data stream, and frameworks like Geant4, Gaudi and Art need to be retrofitted to accommodate the possibility to correlate information across separate events. Moreover, given the challenges involved in updating all custom code built around a specific framework release, the framework versions rarely get updated, effectively barring the dark matter community from recent computational innovations, such as support for multi-treading/multi-processing and the integration with Graphical Processing Units (GPUs).

Supporting scalable software frameworks across HEP in an experiment-independent fashion will therefore be crucial to our community's success in the next decade. Moreover, given the small scale of the software and computing effort in the different experiments, each collaboration would benefit from a more open sharing of data and software tools, by eliminating substantial duplication of effort and exploiting synergies. Sharing of data and resources hasn't been encouraged until now, due to a patchwork of funding agencies (with different data sharing requirements) and competition between various collaborations. Recent efforts from the Liquid Xenon~\cite{Aalbers:2022dzr} and DANCE~\cite{Dance:2022} communities are promising developments in the direction of maximal collaboration and resource sharing.

\subsection{Machine learning techniques}

Recent advances in Machine Learning (ML) have enabled new ways to analyze large scientific datasets. The HEP community has successfully leveraged these advances to improve event energy and position reconstruction and particle identification, as well as fast simulation and triggering~\cite{Albertsson2018-dt,Karagiorgi:2021ngt}. Existing direct detection experiments have demonstrated improvements in background rejection using simple neural networks and boosted decision trees~\cite{CDMS-ML}. Deep learning techniques have also successfully been employed for reconstruction of physical quantities, such as energy and position, particle identification and signal/background discrimination~\cite{Liang:2021nsz}.

Looking beyond these initial successes, using ML to its full potential will require addressing several challenges at the forefront of scientific computing. Core among these is the need to establish ML-based approaches that can match traditional analyses in their reliability and robustness.
Quantitative uncertainty estimation in ML outputs is integral to translating results to a final, rigorous statistical analysis. Additionally, tools to improve interpretability of ML outputs can allow human analyzers to benefit from their insights, re-examining their assumptions about what information is physically relevant. Uncertainty estimation and interpretability are distinctive ML problems in the science discovery domain, and it is desirable that the direct detection community leverage synergies across HEP to tackle them, to ensure that our results are believable going forward.

Algorithms currently used in the field are largely based on techniques established in the private technology sector, while our unique problem space may lend itself to approaches that are new to us, such as graphical models. The development of new algorithms, ideally in collaboration with ML experts outside the field of particle physics, has the potential to significantly improve performance on problems that do not readily map onto established paradigms. Finally, gaining access to state-of-the-art software tools and hardware resources specific to deep learning, such as clusters of graphical and tensor processing units, will help ensure competitiveness with other related fields.

\subsection{Workforce development}
\label{subsec:Workforce}

The general challenge of maintaining software and computing talent is exacerbated in the direct detection community by the lack of long term, permanent positions within the experiments. Most of the computing-centric HEP positions worldwide are supported by CERN for the needs of the LHC experiments, or by FNAL for the needs of DUNE, leaving the direct detection experiments unsupported. Many of the software tasks are carried out by early-career collaborators, which causes significant turnover from year to year.

The recruiting and retention of scientists who are proficient in both physics and computing has proven increasingly challenging, especially with the rise of ``Data Science'' opportunities in industry, for which HEP alumni are uniquely qualified. Competition for top talent is fierce, notably in high-cost-of-living areas, where the scientific career does not always provide attractive levels of prestige and economic opportunity. A salient example is the NESAP program described in~\ref{subsection:CompModel}, which advertises approximately 25 fellowships per cycle, and has a vacancy rate of approximately 1/3 at any time during the cycle.

It is therefore essential to provide funding for permanent software and computing experts. The careers of these researchers should be evaluated appropriately with regards to efficiency, stability, and robustness. One possibility is the creation of research software engineer positions that have long term funding independent of experiments, but for the support of existing and future experiments. Another area of opportunity is the increase of joint particle physics and data science appointments at universities, which have become marginally more common over the last decade. We also need to increase the general software literacy of physicists, through (continuing) education initiatives and collaborations with industry, national labs and academia. A more transparent approach to software development and data sharing, as outlined in~\ref{subsec:HepFrameworks}, would go a long way towards improving the career prospects of software and computing experts, as it would allow individuals to claim credit for their work and be evaluated appropriately.

The diversity issues pervasive in HEP are exacerbated in the computing domain due to the (perceived) technical nature of the work, and we must ensure that faculty, staff, and trainee positions are viable career options for a diverse group of people. Our workforce development efforts should explicitly include equity, and recognize that diversity is foundational to our success, as it demonstrably increases the creativity of solutions and variety of approaches. Attracting a diverse pool of applicants would help ameliorate our recruiting and retention challenges, as ``success'', ``achievement'' and ``prestige'' are bound to have different meanings for different groups. Diversifying the computational workforce will require efforts to diversify physics in general, and to leverage our partnership with the astronomy and industry communities, who have made substantial efforts in this direction over the last decade. This remains a challenging problem, but not insurmountable, as STEM identity formation for underrepresented minorities has become a priority in a variety of educational settings, from early childhood to higher education~\cite{STEM-diversity}.

\section*{\label{sec::acknowledgments}Acknowledgments}

This work is supported in part by the US Department of Energy, under contract numbers DE-SC0015655 and DE-AC02-76SF00515.

\bibliographystyle{JHEP}
\bibliography{main.bib}

\end{document}